\documentclass[twocolumn,showpacs,preprintnumbers,amsmath]{revtex4}
\usepackage{amssymb}

\usepackage{graphicx}

\tighten 

\begin{document}

\title{Holes as Dipoles in a Doped Antiferromagnet and Stripe Instabilities}

\begin{abstract}
Based on an effective model of a doped antiferromagnetic Mott insulator, we
show that a doped hole will induce a dipole-like spin configuration in a
spin ordered phase at low doping. The kinetic energy of doped holes is
severely frustrated and a hole-dipole object is actually localized or
self-trapped in space. Without a balance from the kinetic energy, the
long-range dipole-dipole interaction between doped holes will dominate the
low-energy physics, leading to an inhomogeneity instability as hole-dipoles
collapse into stripes. Both antiphase metallic stripes of quarter-filling
and antiphase insulating stripes along a diagonal direction are discussed as
composed of hole-dipoles as elementary building blocks. Stripe melting and
competing phases are also discussed.
\end{abstract}

\author{Su-Peng Kou$^{a,b}$ and Zheng-Yu Weng$^a$}

\address{$^a$Center for Advanced Study, Tsinghua University, Beijing 100084\\
$^b$Department of Physics, Beijing Normal University, Beijing, 100875}
\maketitle
\section{Introduction}

The theoretical challenge of doped antiferromagnetic (AF) Mott insulator
systems \cite{anderson} comes from their nature of strong correlations. It
makes the relevant models like the Hubbard and $t-J$ Hamiltonians highly
difficult to deal with. In such systems, the kinetic energy of charge
carriers usually gets strongly frustrated \cite{trugman,string0} and there
is no single \emph{dominant} process based on which a perturbative approach
can be constructed. This opens door for various competing orders at low
doping and makes the phase diagram complex \cite{review1}.

So far there still lacks a single microscopic description which can unify
various competing phenomena known from both experiments in cuprates and
theoretical considerations based on doped antiferromagnets. Given the fact
that exact solutions of two-dimensional (2D) Hubbard and $t-J$ models are
absent, a simplified and tractable model of doped AF Mott insulators is thus
highly desirable in order to have a framework for a systematic study.

Such a type of models should incorporate three most essential ingredients of
a doped AF Mott insulator. Namely, it should properly describes AF
correlations at half-filling, including both\emph{\ long-range} (low-energy)
and \emph{short-range} (high-energy) ones; Its Hilbert space should be
restricted due to the presence of an \emph{upper} Hubbard band at finite
doping; Finally, it should explicitly embody essential frustration effects
introduced by hole hopping in an antiferromagnet. Furthermore, compared to
the original Hubbard and $t-J$ models, it ought to be simplified enough to
allow a tractable treatment.

In this paper, we study an effective model \cite{string2,string1} of a doped
AF-Mott insulator which satisfies the above criteria. This model well
describes AF correlations at half-filling and precisely incorporates the
so-called phase string effect at a finite doping. The latter has been
identified \cite{string0} as a crucial dynamic frustration effect introduced
by holes in the $t-J$ Hamiltonian. Here a phase string depicts a string of
displaced \emph{transverse }spins along a hole-hopping path, characterized
by a sequence of signs: 
\begin{equation}
(+1)\times (-1)\times (-1)\times \text{ }\cdot \cdot \cdot  \label{string}
\end{equation}
as illustrated by Fig. 1(a)$.$ It can be rigorously shown \cite{string0}
that such transverse spin mismatches cannot be self-healed simultaneously
with the displaced spins in the \emph{longitudinal} (quantization) direction
[see Fig. 1(b)], through the superexchange term of the $t-J$ Hamiltonian.
Consequently, after a hole moves through a closed path back to its origin
and the displaced spins by hole hopping are restored in the quantization
direction ($\hat{z}$-axis), the wavefunction should always pick up an
additional Berry-phase-like sign given by (\ref{string}), which is highly
nontrivial in a general case.

\begin{figure}[tbp]
\begin{center}
\includegraphics{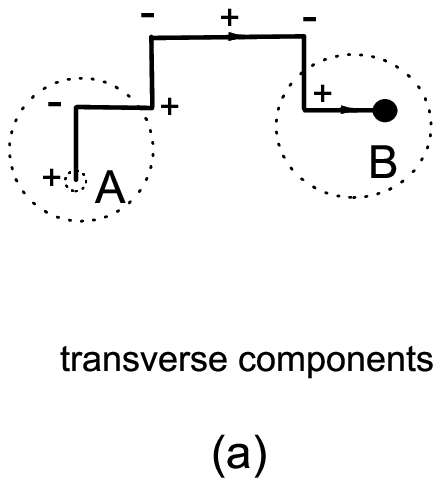} \includegraphics{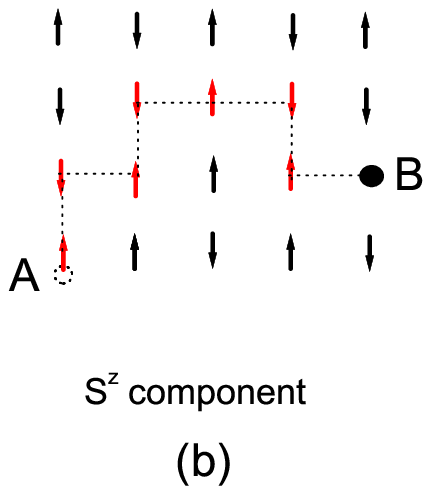}
\end{center}
\caption{Spin mismatches created by a hole hopping in a N\'{e}el order: (a)
in transverse directions (phase string) and (b) in longitudinal direction.
Two dotted circles in (a) indicate the sigularity of the starting and ending
points in a holon path (see text). }
\label{fig1}
\end{figure}

Such a working model may be called the phase string model (section II).
Based on it, we will show that in a spin ordered phase at low doping, a
doped hole will generally induce a dipole-like spin configuration (section
III). The kinetic energy of doped holes is severely frustrated by the phase
string effect such that a hole-dipole is localized or self-trapped in space
with an infinite mass. Without the balance from the kinetic energy, the
long-range dipole-dipole interaction between doped holes will dominate the
low-energy physics. It will lead to an inhomogeneous instability as
hole-dipoles collapse into stripes (section IV). These anti-phase stripes
can be understood as that holes are threaded together by the phase strings.
Both metallic stripes of anti-phase and quarter-filling and insulating
stripe along the diagonal direction are discussed based on the elementary
hole-dipoles. The stripe phase at low doping is highly competitive to \emph{%
homogeneous phases} previously studied \cite{kou} in the same model, which
include the AF long-range ordered state, spin glass state, and
superconducting state. Therefore, a rich array of phases are revealed in the
phase string model, where AF correlations and phase string effect serve as
basic driving forces.

\section{An effective model of a doped antiferromagnet}

The $t-J$ Hamiltonian is one of the simplest nontrivial models of doped Mott
insulator/doped antiferromagnets \cite{anderson}. Based on the exact phase
string formulation \cite{string2} of the $t-J$ Hamiltonian, an effective
low-energy model can be derived, known as the phase string model \cite
{string1}.

The presence of short-range AF correlations provides a natural
``ultraviolet'' cutoff in deducing such an effective model from the $t-J$
Hamiltonian. So the local AF correlations, characterized by the \emph{bosonic%
} RVB pairing \cite{lda}, underpin this effective theory. It differs from
the slave-boson mean-field theories \cite{fRVB} by emphasizing the
underlying AF correlations; It differs from the slave-fermion mean-field
theories \cite{aa,aa1} by precisely dealing with a \emph{singular} hopping
effect in the $t-J$ model, i.e., the phase string effect mentioned in
Introduction.

Such an effective model itself defines a class of doped Mott-insulator/doped
antiferromagnet. It has been shown to possess the following properties: (I)
It well describes antiferromagnetism at half-filling, including both
long-wavelength, low-energy and short-range, high-energy (high-temperature)
physics; (III) It has a d-wave superconducting ground state beyond a finite
doping concentration; (II) It characterizes the correct restricted Hilbert
space for charge and spin degrees of freedom of a doped Mott insulator, when
holes are introduced.

The effective Hamiltonian can be written \cite{string2,string1} as $%
H_{string}=H_h+H_s,$ where 
\begin{eqnarray}
H_h &=&-t_h\sum_{\langle ij\rangle }\left( e^{iA_{ij}^s-i\phi
_{ij}^0}\right) h_i^{\dagger }h_j+H.c.  \label{hh} \\
H_s &=&-J_s\sum_{\langle ij\rangle \sigma }\left( e^{i\sigma
A_{ij}^h}\right) b_{i\sigma }^{\dagger }b_{j-\sigma }^{\dagger }+H.c.
\label{hs}
\end{eqnarray}
with $h_i^{\dagger }$ and $b_{i\sigma }$ denoting \emph{bosonic} holon and
spinon operators, respectively. When the holon number is zero, $H_s$ is
equivalent to the Schwinger-boson mean-field Hamiltonian \cite{aa}, which
well describes AF correlations at half-filling. The AF long range order is
realized by the spinon Bose condensation, i.e., $\left\langle b_{i\sigma
}\right\rangle \neq 0$ at zero temperature. At larger doping, when holons
are Bose condensed, the ground state becomes a (d-wave) superconductor \cite
{muthu}.

A unique feature of this effective model is the presence of link fields in
Eqs. (\ref{hh}) and (\ref{hs}). They are not free gauge fields as they
satisfy the topological constraints 
\begin{equation}
\sum_cA_{ij}^h=\pm \pi \sum_{l\in c}n_l^h,
\end{equation}
and 
\begin{equation}
\sum_cA_{ij}^s=\pm \pi \sum_{l\in c}(n_{l\uparrow }^b-n_{l\downarrow }^b),
\end{equation}
for a closed loop $c$, in which $n_l^h$ and $n_{l\alpha }^b$ are holon and
spinon number operators, respectively. In addition to $A_{ij}^s,$ $\phi
_{ij}^0$ in Eq. (\ref{hh}) describes a uniform $\pi $ flux per plaquette ($%
\sum_{\square }\phi _{ij}^0=\pm \pi $). This model has a \textrm{U(1)}$%
\times $\textrm{U(1)} gauge symmetry as $H_h$ and $H_s$ are invariant under
the gauge transformations: 
\begin{equation}
h_j\rightarrow h_j\exp (i\varphi _j),\text{ }A_{ij}^s\rightarrow
A_{ij}^s+i(\varphi _i-\varphi _j),
\end{equation}
and 
\begin{equation}
b_{j\sigma }\rightarrow b_{j\sigma }\exp (i\sigma \phi _j),\text{ }%
A_{ij}^h\rightarrow A_{ij}^h+i(\phi _i-\phi _j).
\end{equation}

Here the holons and spinons are \emph{basic building blocks} that properly
characterize the restricted Hilbert space of a doped Mott insulator. But it
is important to note that they, as independent entities, are not necessarily
the true low-lying elementary excitations of the system. In this model, the
electron operator is made of holon and spinon operators in the following
form \cite{string2} 
\begin{equation}
c_{i\sigma }=h_i^{\dagger }b_{i\sigma }e^{i\hat{\Theta}_{i\sigma }},
\label{mutual}
\end{equation}
where $\hat{\Theta}_{i\sigma }$ is a topological phase, which ensures the
fermionic statistics of $c_{i\sigma },$ as defined by 
\begin{equation}
e^{i\hat{\Theta}_{i\sigma }}=(-\sigma )^ie^{\frac i2\left[ \Phi _i^b-\sigma
\Phi _i^h\right] },  \label{theta}
\end{equation}
with 
\begin{equation}
\Phi _i^b=\sum_{l\neq i}\mathop{\rm Im}\ln (z_i-z_l)\left( \sum_\alpha
\alpha n_{l\alpha }^b-1\right) ,~  \label{phib}
\end{equation}
and 
\begin{equation}
\Phi _i^h=\sum_{l\neq i}\mathop{\rm Im}\ln (z_i-z_l)n_l^h~.  \label{phih}
\end{equation}
Thus, going from the holon-spinon representation to the electron
representation, topological (vortexlike) phases, $\Phi _i^b$ and $\Phi _i^h,$
are involved, which implies that the holon and spinon objects defined in the
phase string model are topological entities themselves, as will soon become
clear below.

\section{Doped holes: Dipoles in spin ordered phase}

How doped holes propagate in an antiferromagnet and how background spins are
influenced by the hole motion are two central issues in the study of doped
antiferromagnets. In the present section, we will focus on these issues in a
spin ordered phase at small doping, based on the above phase string model.

\subsection{Holons as merons}

The spin-flip operator can be expressed by 
\begin{equation}
S_i^{+}=(-1)^ib_{i\uparrow }^{\dagger }b_{i\downarrow }\exp \left[ i\Phi
_i^h\right] ,  \label{s+^}
\end{equation}
according to Eq. (\ref{mutual}). The presence of the vortex phase $\Phi _i^h$
in Eq. (\ref{s+^}) clearly shows that a holon is a topological object that
affects spins nonlocally.

At half-filling where $\Phi _i^h=0,$ the Bose condensation $<b_{i\sigma
}>\neq 0$ of spinons gives rise to an AF long-range order of spins lying in
the x-y plane: $\left\langle S_i^{+}\right\rangle =(-1)^i<b_{i\uparrow
}^{\dagger }><b_{i\downarrow }>$. Through $\Phi _i^h,$ each holon will
induce a ``meron'' spin twist, which holds true as long as $<b_{i\sigma
}>\neq 0.$ Such a spin vortex configuration (for a holon at the origin) may
be characterized by the unit vector 
\[
\mathbf{n}_i\propto \mathbf{(-}1\mathbf{)}^i\left\langle \mathbf{S}%
_i\right\rangle , 
\]
as follows: 
\begin{equation}
\mathbf{n}_i\mathbf{=r}_i/\mid \mathbf{r}_i\mid ,\text{ \ \ }
\end{equation}
with $\mathbf{r}_i^2=x_i^2+y_i^2.$ It is schematically shown in Fig. 2.

\begin{figure}[tbp]
\begin{center}
\includegraphics{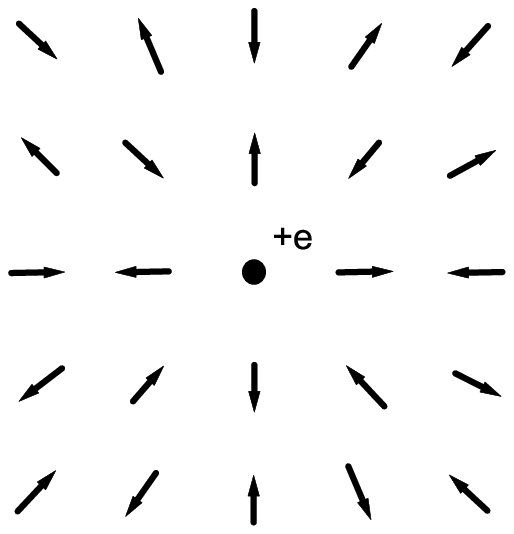}
\end{center}
\caption{A meron-like spin configuration around a holon. }
\label{fig2}
\end{figure}

To evaluate the potential energy of a holon (meron), we consider the spinon
condensed part in $H_s$ [Eq. (\ref{hs})]: 
\begin{equation}
\Delta E_s\simeq -J_s\rho _c^sa^2\sum_{\left\langle ij\right\rangle }\left[
\cos A_{ij}^h-1\right] +c.c.  \label{es1}
\end{equation}
in which $\rho _c^sa^2\equiv <b^{\dagger }>^2$($a$ is the lattice constant)
and a reference energy corresponding to $A_{ij}^h=0$ has been subtracted.
Explicitly $A_{ij}^h$ reads

\begin{equation}
A_{ij}^h=\frac 12\sum_{l\neq i,j}\mathop{\rm Im}\ln \frac{z_i-z_l}{z_j-z_l}%
n_l^h,  \label{ah1}
\end{equation}
and then one has 
\begin{eqnarray}
\Delta E_s &\simeq &J_s\rho _c^sa^2\sum_{\left\langle ij\right\rangle
}\left( A_{ij}^h\right) ^2  \nonumber \\
&\equiv &E_s^m+E_s^{m-m},
\end{eqnarray}
where $E_s^m=\sum_ln_l^h\varepsilon ^m$ with a single holon-meron
self-energy 
\begin{eqnarray}
\varepsilon ^m &\equiv &\frac{J_s\rho _c^sa^2}4\sum_{\left\langle
ij\right\rangle \left( \neq l\right) }\left( \mathop{\rm Im}\ln \frac{z_i-z_l%
}{z_j-z_l}\right) ^2  \nonumber \\
&\simeq &\frac \pi 2J_s\rho _c^sa^2\mathrm{\ln }({L/a),}  \label{em}
\end{eqnarray}
where $L$ is the size of the sample. Namely, a single holon as a meron in
the spinon condensed phase will generally cost a logarithmically divergent
energy.

Furthermore, the meron-meron interaction is given by 
\begin{equation}
E_s^{m-m}=\sum_{l<l^{^{\prime }}}n_l^hV_{ll^{^{\prime }}}^{m-m}n_l^h,
\label{int}
\end{equation}
and 
\begin{eqnarray}
V_{ll^{^{\prime }}}^{m-m} &=&\frac{J_s\rho _c^sa^2}2\sum_{\left\langle
ij\right\rangle \left( \neq ll^{^{\prime }}\right) }\mathop{\rm Im}\ln \frac{%
z_i-z_l}{z_j-z_l}\mathop{\rm Im}\ln \frac{z_i-z_{l^{^{\prime }}}}{%
z_j-z_{l^{^{\prime }}}}  \nonumber \\
&\simeq &-\pi J_s\rho _c^sa^2\ln \frac{\left| \mathbf{r}_l-\mathbf{r}%
_{l^{^{\prime }}}\right| }L.  \label{v}
\end{eqnarray}
Here two holons will always \emph{repulse} each other logarithmically, as
they are merons of the \emph{same} topological charge.

We note that there have been proposals of hole-merons in literature \cite
{meron,timm}. They usually involve both \emph{merons} and \emph{anti-merons}
. Two holes may thus be bound like a meron-anti-meron (vortex-antivortex)
pair. By contrast, in the present case holons are always \emph{repulsive} to
each other as they carry the same topological charge. The underlying
physical reason is that this is in favor of the kinetic energy for a finite
concentration of holons, which become identical bosons after explicitly
incorporating the phase string effect in the slave-fermion formalism.
Consequently, superconductivity in the present doped AF-Mott insulator
originates \cite{muthu} from the \emph{Bose condensation} of holons, instead
of the \emph{pairing} of them. Here one should distinguish the \emph{holon
pairing} from the pairing order parameter of \emph{electrons. }The latter
still characterizes the superconducting state in the present case where
spins are in RVB pairing, even though the holon pairing is absent.

Intuitively one might envisage that a phase string induced by one hole be
``erased'' by an another hole accompanying it. This may look like in favor
of hole pairing. But such a picture is actually more in favor of stripes in
which many holes precisely follow each other's paths to eliminate the phase
string effect. The basic reason is that the phase string effect of (\ref
{string}) is not simply a ``geometric'' effect: it also involves spin
dynamics and is very \emph{singular}. Thus, unless one hole\emph{\ tightly}
follows the other along the \emph{same} path, like in the stripe case, the
singular dynamic phase string effect cannot be completely removed by two
holes simply staying close to each other. In fact, previously Trugman \cite
{trugman,review1} has already found the presence of frustrations for a hole
pair's motion in an antiferromagnet, due to \emph{a quantum effect}
originating from the fermionic character of the background spins. This
effect is actually included in the phase string effect in the present
approach. Finally, we note that technically, a holon is just a building
block in the phase string model and the possibility that some holons may be
anti-merons can be, in principle, constructed as composite objects in the
present framework. But such a construction is usually not energetically
favorable at finite doping and will be discussed elsewhere. The pairing of
two doped holes has been discussed \cite{eder} in the $t-J_z$ model based on
the ``string'' picture before.

\begin{figure}[tbp]
\begin{center}
\includegraphics{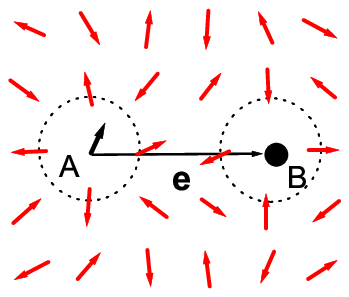}
\end{center}
\caption{A hole-dipole is a confined object composed of a holon-meron and an
anti-meron at two poles: B and A, connected by a branch-cut with the spatial
separation $\mathbf{e}$ as a dipole moment. }
\label{fig3}
\end{figure}

\subsection{Holon confinement: Doped holes as dipoles}

To remove such an energy divergence [Eq.(\ref{em})] associated with a meron
spin configuration, each holon will have to ``nucleate'' an anti-meron from
the vacuum. Being bound to the latter, a dipole composite with a finite
energy will be formed. This elementary object will carry a charge $+e$ and
replace the holon to represent a true stable \emph{hole} in the spin ordered
phase. By contrast, free \emph{holons} disappear from the finite-energy
spectrum. We can also call this phase as the holon confined phase.

\subsubsection{\emph{Formation of dipoles}}

An anti-meron configuration is defined as a spin twist 
\begin{equation}
b_{i\sigma }\rightarrow \tilde{b}_{i\sigma }\exp \left[ i\frac \sigma
2\vartheta _i^k\right]  \label{bmeron}
\end{equation}
where $\vartheta _i^k=\mathop{\rm Im}\ln (z_i-z_k^0).$ Here $z_k^0$ denotes
the position (center) of an anti-meron labeled by $k$. Correspondingly 
\begin{equation}
\left\langle S_i^{+}\right\rangle \rightarrow (-1)^i\left\langle \tilde{b}%
_{i\uparrow }^{\dagger }\right\rangle \left\langle \tilde{b}_{i\downarrow
}\right\rangle \exp \left[ i\Phi _i^h-i\Omega _i\right]  \label{s+}
\end{equation}
where $\Omega _i=\sum\nolimits_k\vartheta _i^k$ (summing over all induced
anti-merons). If $z_k^0$'s are close to the positions of corresponding
holons, then cancellation occurs in $\Phi _i^h-\Omega _i.$ Consequently, the
logarithmic divergence is removed from $H_s$ (see below).

Define $n_i^x+in_i^y=e^{i\phi _0+i\phi _i}$, with the unit vector $\mathbf{n}%
_0\equiv $ $(\cos \phi _0,\sin \phi _0)$ as the magnetization direction at
infinity$.$ For a bound pair of meron (holon)-anti-meron centered at the
origin, labeled by the index $k$, one obtains according to (\ref{s+})

\begin{equation}
\phi _i^k=\text{Im }\ln \text{ }\frac{z_i-z_k/2}{z_i+z_k/2},  \label{phik}
\end{equation}
with $z_k\equiv e_k^x+ie_k^y.$ Here $\mathbf{e}_k$ denotes the spatial
displacement of the meron and anti-meron centered at $\pm \frac{\mathbf{e}_k}%
2,$ respectively. At $|\mathbf{r}_i|>>|\mathbf{e}_k\mathbf{|,}$ 
\begin{equation}
\phi _i^k\simeq \frac{\left( \mathbf{\hat{z}\times e}_k\right) \mathbf{\cdot
r}_i}{|\mathbf{r}_i|^2}.  \label{dipole}
\end{equation}
Thus, each pair of holons and anti-merons forms a composite, as shown in
Fig. 3, that has a \emph{dipolar} spin configuration at a sufficiently large
distance.

In $H_s$, one may easily check that $\omega _{ij}\equiv (\Omega _i-\Omega
_j)/2$ cancels out $A_{ij}^h$ and thus removes the aforementioned energy
divergence. Here the energy cost Eq. (\ref{es1}) is replaced by 
\begin{eqnarray}
\Delta E_s &=&-2J_s\rho _c^sa^2\sum_{\left\langle ij\right\rangle }\left[
\cos \left( A_{ij}^h-\omega _{ij}\right) -1\right]  \nonumber \\
&\simeq &J_s\rho _c^sa^2\sum_{\left\langle ij\right\rangle }\left(
A_{ij}^h-\omega _{ij}\right) ^2  \nonumber \\
&\equiv &\sum_k\mathcal{E}_k^d+\sum_{k<k^{^{\prime }}}V_{kk^{\prime }}^{d-d},
\label{es}
\end{eqnarray}
in which the single dipole energy 
\begin{equation}
\mathcal{E}_k^d=\frac{J_s\rho _c^sa^2}4\sum_{\left\langle ij\right\rangle
}\left( \phi _i^k-\phi _j^k\right) ^2,
\end{equation}
and dipole-dipole interaction energy 
\begin{equation}
V_{kk^{\prime }}^{d-d}=\frac{J_s\rho _c^sa^2}2\sum_{\left\langle
ij\right\rangle }\left( \phi _i^k-\phi _j^k\right) \left( \phi _i^{k^{\prime
}}-\phi _j^{k^{\prime }}\right) .  \label{dd}
\end{equation}
In the continuum limit, $\phi _i^k-\phi _j^k\approx (\mathbf{r}_i-\mathbf{r}%
_j)\cdot \nabla \phi ^k$ and, according to Eq. (\ref{phik}),

\begin{equation}
\nabla \phi ^k=\frac{\mathbf{\hat{z}\times }\left( \mathbf{r}-\mathbf{e}%
_k/2\right) }{\left| \mathbf{r}-\mathbf{e}_k/2\right| ^2}-\frac{\mathbf{\hat{%
z}\times }\left( \mathbf{r}+\mathbf{e}_k/2\right) }{\left| \mathbf{r}+%
\mathbf{e}_k/2\right| ^2}.  \label{gphi}
\end{equation}
Then the self-energy of a dipole can be evaluated by $\mathcal{E}_k^d\simeq 
\frac{J_s\rho _c^sa^2}4\int d^2\mathbf{r}\left( \nabla \phi ^k\right) ^2.$
In contrast to the logarithmically divergent meron-energy$,$ there is a
branch-cut connecting two poles at $\pm \frac{\mathbf{e}_k}2$\ of the hole
dipole, and $\mathcal{E}_k^d$ becomes finite 
\begin{eqnarray}
\mathcal{E}_k^d &\simeq &\frac{J_s\rho _c^sa^2}4\int d^2\mathbf{r}\frac{%
\left| \mathbf{e}_k\right| ^2}{\left| \mathbf{r}-\mathbf{e}_k/2\right|
^2\left| \mathbf{r}+\mathbf{e}_k/2\right| ^2}\text{ }  \nonumber \\
&\simeq &\pi J_s\rho _c^sa^2\ln \frac{\left| \mathbf{e}_k\right| +a}a,\text{%
\qquad ~}  \label{edipole}
\end{eqnarray}
at $\left| \mathbf{e}_k\right| \gtrsim a$.

\subsubsection{\emph{Type A and Type B dipoles}}

Let us first point out that generally there are two possible anti-meron
configurations as defined by Eq. (\ref{bmeron}). Type A: the core sits on a
lattice site $l$, i.e., $z_k^0=z_l.$ In this case, there will be always a
spin 1/2 sitting at the core of the anti-meron [Fig. 3], according to the
single-occupancy constraint. Type B: $z_k^0$ is located inside a lattice
plaquette, instead of a lattice site. Figs. 4(a) and (b) illustrates both
cases at minimal sizes.

\begin{figure}[tbp]
\begin{center}
\includegraphics{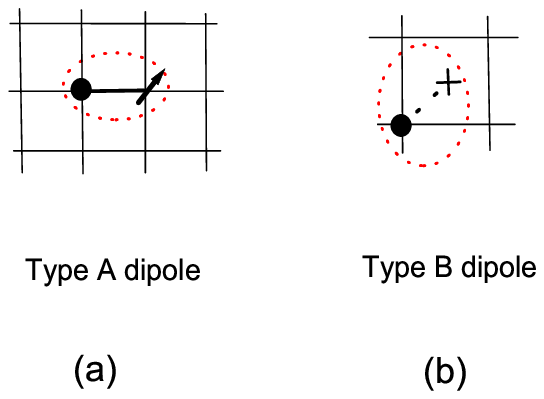}
\end{center}
\caption{Two typical hole-dipoles of minimal size: (a) Type A dipole with a
holon and a spin sitting at two poles at nearest-neighboring lattice sites;
(b) Type B dipole with one pole at the center of a plaquette.}
\label{fig4}
\end{figure}

\emph{Type A dipole. }Besides the energy consideration given above, it is
important to make sure the self-consistency of the theory, that is, the
anti-meron configuration Eq. (\ref{bmeron}) should not violate the \emph{%
single-valueness} of the electron operator (the spin operator in Eq. (\ref
{s+}) is obviously single-valued).

Consider that a winding of the coordinate $z_i$ around $z_k^0$ which is
located at a lattice site. A phase $\pm \pi $ (corresponding to $\sigma =\pm
1$) will be acquired on the r.h.s. of Eq. (\ref{bmeron}). In order to ensure
the single-valueness of $c_{i\sigma }$, the holon field has to be
transformed in response to Eq. (\ref{bmeron}). We may introduce

\begin{equation}
h_i^{\dagger }\rightarrow h_i^{\dagger }e^{-i\frac{\alpha _k}2\mathop{\rm Im}
\ln \text{ }(z_i-z_k^0)},  \label{hmeron}
\end{equation}
with the spin index $\alpha _k=\pm 1$ denoting the spin at the site $z_k^0.$
Then the total change in the electron operator $c_{i\sigma }$ (for a winding
of the coordinate $z_i$ around $z_k^0)$ will be $e^{i(\sigma \pi -\alpha
_k\pi )}=$ $1$, still satisfying the single-valueness condition.

The physical implication of (\ref{hmeron}) is as follows. It will change $%
H_h $ in Eq. (\ref{hh}) by 
\begin{equation}
A_{ij}^s\rightarrow \tilde{A}_{ij}^s\equiv A_{ij}^s-\frac{\alpha _k}2%
\mathop{\rm Im }\ln \frac{z_i-z_k^0}{z_j-z_k^0}.
\end{equation}
It simply removes the contribution of the spinon, $\sum_\sigma \sigma
n_{l\sigma }^b\equiv \alpha _k,$ at the site of $z_l=z_k^0$ from $A_{ij}^s,$
i.e., 
\begin{equation}
\sum_c\tilde{A}_{ij}^s=\sum_cA_{ij}^s-\alpha _k\pi ,
\end{equation}
where $c$ is a loop encircling $z_k^0.$ By incorporating the phase on the
r.h.s. of Eq. (\ref{hmeron}) into $\hat{\Theta}_{i\sigma }$ of Eq. (\ref
{mutual})$,$ one may also see that the contribution from the spinon at the
site of $z_k^0$ is eliminated from $\Phi _i^b$ and thus from $c_{i\sigma }.$
It means that the spin at $z_k^0$ will no longer exert the \emph{nonlocal
effect, }embedded in a spinon, through $A_{ij}^s$ in $H_h$ and $\hat{\Theta}%
_{i\sigma }$ in $c_{i\sigma }$. Hence\emph{, to maintain the
single-valueness condition after the creation of an anti-meron, the spin 1/2
at its center site }$z_k^0$\emph{\ will lose its spinon character and looks
like a local ``defect'' spin.}

Such a ``defect'' spin, sitting at the core of an anti-meron, will be
decoupled from the environmental spins and simply behave like a free $S=1/2$
spin. This can be straightforwardly checked. By noting that the
transformation Eq. (\ref{bmeron}) excludes the case $z_i=z_k^0$ due to the
uncertainty of $\vartheta _i^k,$ the phases of those terms in $H_s$ that
involve the links attached to the site $z_k^0,$ will become uncertain after
the transformation. Specifically, these terms connected to $z_k^0$ will all
vanish when one averages over $\vartheta _0$ after the replacement $%
\vartheta _i^k\rightarrow \vartheta _i^k+\vartheta _0$ in Eq. (\ref{bmeron})
for $z_i\neq z_k^0$, while the rest of $H_s$ remains invariant. Thus, the
spin 1/2 at $z_k^0$ is indeed decoupled from the environmental spins.

\emph{Type B dipole.} Now consider the case that $z_k^0$ is located at a
center of a plaquette [see Fig. 4(b)]. There is no longer a lattice spin at $%
z_k^0$ and the holon transformation Eq. (\ref{hmeron}) may be replaced by
the form

\begin{equation}
h_i^{\dagger }\rightarrow h_i^{\dagger }e^{\pm i\frac 12\vartheta _i^k}.
\label{hmeron1}
\end{equation}
One can check that $c_{i\sigma }$ still remains single-valued for a winding
of the coordinate $z_i$ around $z_k^0.$ Based on Eq. (\ref{hmeron1}), the
holon Hamiltonian $H_h$ is changed by $A_{ij}^s\rightarrow \tilde{A}_{ij}^s,$
with 
\begin{equation}
\tilde{A}_{ij}^s\equiv A_{ij}^s\pm \frac 12\mathop{\rm Im} \ln \frac{%
z_i-z_k^0}{z_j-z_k^0}.
\end{equation}
Physically it means that an additional $\pm \pi $ flux penetrates through
the plaquette where $z_k^0$ resides: 
\begin{equation}
\sum_c\tilde{A}_{ij}^s=\sum_cA_{ij}^s\pm \pi ,  \label{as2}
\end{equation}
with $c$ as a loop encircling the given plaquette.

In contrast to Type A, it is not obvious that there is a free spin 1/2
trapped around the plaquette of the core of the anti-meron. In Type B, the
cost of the superexchange energy is mainly due to the spin twist around the
plaquette $z_k^0$ which is less dramatic than in Type A, where the presence
of an isolated spin 1/2 at $z_k^0$ means the loss of the superexchange
energy for four nearest-neighboring links connected to it. But in Type B,
there are additional $\pi $ flux quanta [see Eq. (\ref{as2})]\ threading
through the plaquettes, where the cores of anti-merons are located, seen by
holons. So the hopping energy of holons in Type B is different from that of
a Type A dipole. Generally, Type A dipoles are expected to be more stable at
higher doping, while Type B dipoles should be energetically favorable at low
doping.

\subsubsection{\emph{Localization of hole dipoles and phase string effect }}

The fact that a doped hole will induce a dipolar spin configuration was
first pointed out by Shraiman and Siggia \cite{ss} based on a semi-classical
treatment of the $t-J$ model. In Shraiman-Siggia's picture, the hole is
sitting at the \emph{center} of the dipole and such a hole-dipole is mobile
in space. This mobility may be easily understood if one considers the
translational invariance and the fact that the hopping overlap integral for
two hole-dipoles with their centers displaced by a lattice constant should
be finite in general. That a mobile hole carries a long-range dipolar spin
distortion has been also studied in Refs. \cite{frenkel,manousakis}.

By contrast, in the present case, the hole is sitting at \emph{one} of the
two opposite poles of a dipole, \emph{not} at its center. The holon itself
has a \emph{finite} hopping integral in $H_h$ and can hop around.
Nonetheless it is confined to an anti-meron by a logarithmic potential in
Eq. (\ref{edipole}). On the other hand, an anti-meron itself as a
semi-classical vortex formed by the condensed spinons is immobile: it cannot
propagate in terms of $H_s.$ In other words, its effective mass is infinite,
which is in contrast to a holon which has a finite hopping integral, or
mass, according to $H_h$. Therefore, the hole dipole (both Type A and B) as
a whole must remain localized or self-trapped in space.

This distinction between the Shraiman-Sigga dipole and the present one can
be traced back to a very singular effect in the $t-J$ model, which has been
inappropriately treated in the semi-classical approach of Shraiman-Sigga. In
the following we further clarify this issue.

According to Eq. (\ref{phik}), there is a branch-cut connecting two poles
(at $\pm \frac{\mathbf{e}_k}2,$ respectively$)$ of a hole dipole at the
origin, as illustrated in Fig. 3. Such a branch-cut can be physically
understood in terms of the irreparable phase string effect, which was
discovered \cite{string0} as a generic property of the $t-J$ model.

A phase string is basically a sequence of $+$ and $-$ signs shown in (\ref
{string}) [also see Fig. 1(a)], which is left by a spinless hole on the path
as it moves on a spin background described by the $t-J$ model. The origin of
the phase string is due to the disordered Marshall sign caused by the hole
hopping \cite{string0}. Physically it may be interpreted as describing
displaced spins, by hopping, along the\emph{\ transverse} directions (in x-y
plane). Specifically, the sign $+$ and $-$ in (\ref{string}) distinguish the
``backflow'' spin $\uparrow $ and $\downarrow $ exchanged with a hole at
each step of hopping. Since the superexchange term ``respects'' the Marshall
sign, the phase string, describing the displaced Marshall sign, cannot be
``repaired'' through the superexchange term. Therefore, if one draws a
circle around either the hole at B or the starting point at A, it will
always intercept the phase string once (or odd times) [see Fig. 1(a)].

Pictorially, one may consider a bare hole created by the electron $c$%
-operator at a point A and then imagine that the spinless hole (holon) hops
to the point B. It results in a dipole configuration with the two poles as
its starting point A and ending point B, connected by the phase string in
between [Fig. 1(a)]. Since the holon can reach B from different paths
originated at A, the singular phase string is then replaced by (or relaxes
to) a smoother \emph{dipole} configuration. Nonetheless, the topology
remains as represented by the branch-cut connecting two poles of the dipole
as illustrated by Fig. 3.

In a N\'{e}el state with the magnetization lying in x-y plane, the
displacement of staggered magnetization moments will reach the maximum,
namely, $\pi $ phase per site (inversion of the spin polarization) on the
hole path. It implies that going through the circle once, around the point A
or B in Fig. 1(a), one finds the total spin displacement relative to the
original N\'{e}el spin background is $\pi $, modulo $2\pi $, which has the
same topology as a meron or anti-meron discussed above.

Finally, the localization of a hole dipole can be physically understood in
terms of the phase string picture too \cite{1hole}. Since the phase string (%
\ref{string}) cannot be ``self-healed'' through, say, spin dynamics, a holon
moves far away from its origin A will eventually lose its phase coherence as
the product (\ref{string}) becomes uncertain due to the zero-point spin
fluctuations in the N\'{e}el background. It means the propagating amplitude
vanishes such that the holon must be localized or self-trapped around the
origin A within a finite distance. The single electron spectral function for
the one-hole doped case has been studied \cite{1hole} based on such a holon
localization description, in which the dispersion of the low-lying peak is
originated from the ``spinon'' sepectrum instead. The calculated spectral
function well explains the corresponding photoemission experiments \cite
{1hole}.

Delocalization of holons will occur at higher doping when the spin
background is changed (self-consistently) by the phase string effect. Such a
holon deconfined phase is generally a superconducting state as discussed in
Refs. \cite{kou,muthu} On the other hand, at low doping, without the
destruction of spin ordering, delocalization of holons can only be realized
in the reduced dimension when holons follow each other's paths such that the
induced phase string frustrations can be eliminated. Such a metallic stripe
in the spinon condensed phase will be discussed in the following section.

\section{Stripe instability}

As discussed above, a hole-dipole object does not have a kinetic energy, or,
its self-energy does not have a dispersion, since it is self-trapped in
space with an infinite mass. Thus the long-range dipole-dipole interaction
among dipoles may become crucial in determining the low-energy physics in
such a phase where kinetic energy is strongly frustrated. In this part we
will discuss the origin of stripes based on this picture.

\subsection{Dipole--dipole interaction}

For a dipole located at the origin, Eq.(\ref{gphi}) can be reduced to 
\begin{equation}
\nabla \phi ^k\simeq \frac 1{\left| \mathbf{r}\right| ^2}\left[ 2\frac{%
\left( \mathbf{\hat{z}\times r}\right) \left( \mathbf{r}\cdot \mathbf{e}%
_k\right) }{\left| \mathbf{r}\right| ^2}-\mathbf{\hat{z}\times e}_k\right] ,%
\text{ }~
\end{equation}
for $|\mathbf{r}|>>|\mathbf{e}_k\mathbf{|}$ and a singular part 
\begin{equation}
\nabla \phi ^k\simeq -2\pi \left( \mathbf{\hat{z}\times e}_k\right) \mathbf{%
\delta }\left( \mathbf{r}\right) \mathbf{,}
\end{equation}
at $|\mathbf{r}|\sim 0$ ($|\mathbf{e}_k\mathbf{|\rightarrow }0$)$\mathbf{.}$

Then, in terms of (\ref{dd}), the dipole-dipole interaction between $k$-th
and $k^{\prime }$-th dipoles, located at $\mathbf{r}_k$ and $\mathbf{r}%
_{k^{\prime }}$, respectively, which are well-separated in space by $\left| 
\mathbf{r}_{kk^{^{\prime }}}\right| \equiv \left| \mathbf{r}_k-\mathbf{r}%
_{k^{\prime }}\right| \mathbf{\gg }\left| \mathbf{e}_k\right| $, can be
expressed in leading order approximation by 
\begin{eqnarray}
V_{kk^{^{\prime }}}^{d-d} &=&\frac{J_s\rho _c^sa^2}2\int d^2\mathbf{r}\text{ 
}\mathbf{\nabla }\phi ^k\cdot \nabla \phi ^{k^{\prime }}  \nonumber \\
&\simeq &-J_s\rho _c^sa^2\pi \left[ \left( \mathbf{\hat{z}\times e}_k\right)
\cdot \nabla \phi ^{k^{\prime }}\left( \mathbf{r}_k\right) -\left(
k\leftrightarrow k^{\prime }\right) \right]  \nonumber \\
&\simeq &\frac{2\pi J_s\rho _c^sa^2}{\left| \mathbf{r}_{kk^{^{\prime
}}}\right| ^2}\left[ \mathbf{e}_k\cdot \mathbf{e}_{k^{^{\prime }}}-2\frac{(%
\mathbf{e}_k\cdot \mathbf{r}_{kk^{^{\prime }}})(\mathbf{e}_{k^{^{\prime
}}}\cdot \mathbf{r}_{kk^{^{\prime }}})}{\left| \mathbf{r}_{kk^{^{\prime
}}}\right| ^2}\right]  \label{ddi}
\end{eqnarray}
Here $V_{kk^{^{\prime }}}^{d-d}$ has a typical form of magnetic
dipole-dipole interaction, with $\mathbf{e}_k$ denoting the dipole moment.

Consider orientations of dipole moments which minimize the energy of the
dipole-dipole interaction Eq. (\ref{ddi}) at a fixed distance $\left| 
\mathbf{r}_{kk^{\prime }}\right| $. Define $\cos \varphi _k=\mathbf{e}%
_k\cdot \mathbf{r}_{kk^{\prime }}/\left| \mathbf{e}_k\right| \left| \mathbf{r%
}_{kk^{\prime }}\right| $ and $\cos \varphi _{k^{\prime }}=\mathbf{e}%
_{k^{\prime }}\cdot \mathbf{r}_{kk^{\prime }}/\left| \mathbf{e}_{k^{\prime
}}\right| \left| \mathbf{r}_{kk^{\prime }}\right| .$ We have 
\begin{equation}
V_{kk^{\prime }}^{d-d}=-\frac{2\pi J_s\rho _c^sa^2\left| \mathbf{e}_k\right|
\left| \mathbf{e}_{k^{\prime }}\right| }{\left| \mathbf{r}_{kk^{\prime
}}\right| ^2}\cos \left( \varphi _k+\varphi _{k^{\prime }}\right) ,
\label{ddpa}
\end{equation}
which is minimized at 
\begin{equation}
\varphi _k=-\varphi _{k^{\prime }}+\text{ modulo }2\pi  \label{optimal}
\end{equation}
with 
\begin{equation}
V_{min}^{d-d}=-\frac{2\pi J_s\rho _c^sa^2\left| \mathbf{e}_k\right| \left| 
\mathbf{e}_{k^{\prime }}\right| }{\left| \mathbf{r}_{kk^{^{\prime }}}\right|
^2}.  \label{ddmin}
\end{equation}
Other orientations of dipole moments always give rise to higher energies for
a given $\left| \mathbf{r}_{kk^{\prime }}\right| .$ For instance, it is easy
to see that for the case of random orientations $<\mathbf{e}_k\cdot \mathbf{e%
}_{k^{\prime }}>=0,$ $\left\langle V_{kk^{\prime }}^{d-d}\right\rangle =0$.

\subsection{Stripe instability}

Due to translational invariance, the center of a hole-dipole can be located
at any spatial point with the same self-energy. Thus, to minimize (\ref
{ddmin}), the distance $\left| \mathbf{r}_{kk^{^{\prime }}}\right| $ between
two dipoles should be continuously shrunk. Without a balance from the
kinetic energy, such collapsing seems an inevitable consequence of the
dipole-dipole interaction. Of course, single hole-dipole picture is only
meaningful at a scale $\left| \mathbf{r}_{kk^{^{\prime }}}\right| \gtrsim
\left| \mathbf{e}_k\right| .$ Below such a scale, one has to take into
account the detailed local physics to determine the hole-spin configuration
and energy.

It is important to note that this dipole-dipole ``attraction'' does not
simply result in a real-space pairing of holes. Due to the lack of kinetic
energy cost, more dipoles would rather like to lump together, with
orientations of dipole moments satisfying Eq. (\ref{optimal}), in order to
lower the potential energy. They can thus either line up to form a stripe or
phase separate into Wigner-crystal-like clusters. In reality, the existence
of long-range Coulomb interaction may generally favor \cite{review1} a
stripe configuration over \emph{large} Wigner-crystal-like clusters.
Moreover, the latter possibility can be also considered as a further
instability of stripes when they are attractive to each other and
subsequently collapse into 2D clusters. Thus one can always focus on single
stripes first, based on the hole-dipole picture. In fact, we shall see below
that, for instance, forming a metallic stripe will lead to a disassociation
of individual dipoles and release additional kinetic energies which can
further stabilize stripes against forming a 2D cluster.

One possible loophole in the above argument is the existence of some other
forces which may be in competition with the dipole-dipole interaction. We
first note that the dipole-dipole interaction could be quite weak. In a
hole-dipole shown in Fig. 3, the holon sitting at one pole is actually not
static: it generally will hop around the pole at A, in terms of $H_h$ [Eq. (%
\ref{hh})], such that $\left\langle \mathbf{e}_k\right\rangle $ may get
strongly suppressed, and so does the dipole-dipole interaction. It is a
matter of competition between the hopping energy of \emph{holons} (which
contributes to the self-energy of a hole-dipole) and the dipole-dipole
interaction in Eq. (\ref{ddpa}) or Eq. (\ref{ddmin}) that self-consistently
determines the actual dipole moment $\left\langle \mathbf{e}_k\right\rangle
. $

In realistic situation, for example, a small amount of disorders can easily
pin down hole-dipoles without the penalty from kinetic energy, such that the
distribution of holes may still remain \emph{uniform} on average. Based on a
homogeneous picture, it has been previously shown \cite{kou} that the AF
long range state will persist over some finite doping concentration if
interlayer spin-spin coupling is considered in the same model. The
dipole-dipole interaction will then be responsible for causing a cluster
spin glass phase beyond some critical concentration of holes and eventually
an insulating-superconducting transition takes place at a critical doping $%
x_c\sim 0.043$ \cite{kou}.

Even as a metastable state with local minima, such uniform phases may only
evolve into stripe states of lower energy after a macroscopic time scale. So
these \emph{uniform} insulating phases are highly competitive with the \emph{%
inhomogeneous} phases, in a real experimental situation where a quenched
disordered phase from high temperature may result in a spin glass phase \cite
{kou,aharony,glazman,che,good1}, with the help of impurities, instead of a
more ordered stripe phase at low temperature.

\subsection{Static stripes}

So the long-range dipole-dipole interaction will be responsible for doped
holes to ``talk'' to each other and for them to \emph{collectively} collapse
into stripes. It provides a mechanism for the instability of a homogeneity
insulating phase and the formation of stripes. But the local symmetry and
hole-spin dynamics will ultimately decide which kind of stripes is most
stable. In the following, we consider some general properties of stripes
with hole-dipoles as elementary building blocks.

\emph{Anti-phase. }Let us consider a line-up of hole-dipoles along, say, $%
\hat{x}$-axis with the dipole orientations satisfying the optimal condition
Eq. (\ref{optimal}). Namely, $\left( \mathbf{e}_k\right) _x\equiv e_x,$ $%
\left( \mathbf{e}_k\right) _y\equiv (-1)^ke_y$, where integer $k$ is an
index labelling dipoles. Here we assume that each dipole has the same size
of a dipole moment $\left| \mathbf{e}\right| $ with a minimal spacing $2e_x$
between two dipoles. For a site far away from the $\hat{x}$-axis, the spin
twist is then given by 
\begin{eqnarray*}
\phi _i &=&\sum_k\phi _i^k=\sum_k\frac{e_xy_{ik}-(-1)^ke_yx_{ik}}{r_{ik}^2}
\\
&\simeq &\sum_k\frac{e_xy_{ik}}{r_{ik}^2}=e_x\sum_{k=-\infty }^{+\infty }%
\frac{y_i}{(2e_xk+x_0)^2+y_i^2} \\
&\simeq &\frac \pi 2sgn(y_i),\text{ \qquad }
\end{eqnarray*}
when $\left| y_i\right| \gg \left| \mathbf{e}\right| \sim a.$ Here $%
y_{ik}=y_i$ and $x_{ik}=2e_xk+x_0$ according to the definition.

Thus, an \emph{anti-phase} ($\pi $ phase shift) is found across the stripe
with 
\begin{equation}
\Delta \phi =\phi _{y>0}-\phi _{y<0}=\pi .  \label{deltaphi}
\end{equation}
Namely, a stripe composed of hole-dipoles is topologically always an
anti-phase domain wall of spins.

\emph{Metallic stripe. }For Type A dipoles discussed in Sec. III, a minimal
choice of $\mathbf{e}$ is $e_x=a$ and $e_y=0,$ as shown in Fig. 4(a). If one
simply line up these dipoles along the $\hat{x}$-axis, then the
corresponding stripe is an anti-phase domain wall with quarter-filled holons 
\begin{equation}
n_h=\frac 14  \label{qf}
\end{equation}
(i.e., one hole per two sites), as illustrated in Fig. 5(a).

\begin{figure}[tbp]
\begin{center}
\includegraphics{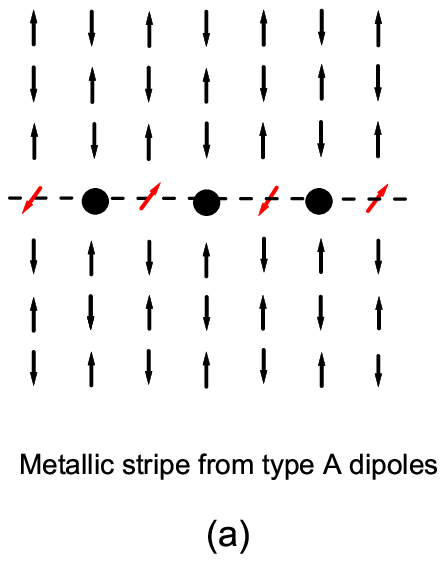} \includegraphics{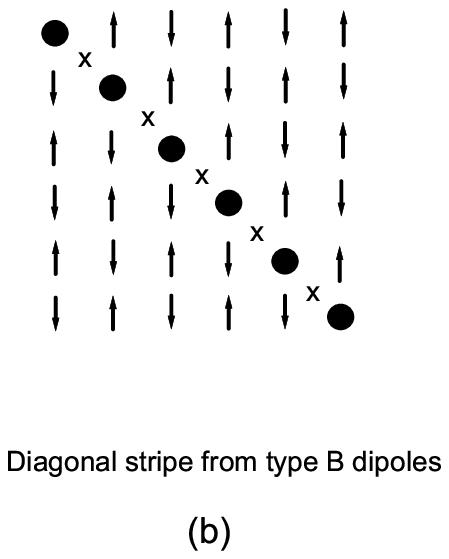}
\end{center}
\caption{Two kinds of stripes as composed of two elementary types of
hole-dipoles shown in Fig. 4: (a) An anti-phase, quarter-filling, metallic
stripe; (b) An anti-phase, insulating stripe.}
\label{fig5}
\end{figure}

Obviously the twists of spins due to the presence of hole-dipoles are all
canceled out away from the stripe, such that the total potential energy is
minimized. Without spin frustrations, two stripes will not gain additional
energy by being close together and thus these stripes are also stable
against a cluster formation. Furthermore, within each anti-phase stripe, the
hole-dipoles will disassociate to gain additional kinetic energy. It is
straightforward to see now holons can move along the stripe without causing
additional frustrations, just like in the one-dimensional (1D) $t-J$ model 
\cite{string2}. Here the flowing of the ``river of charges'' along the
stripe will \emph{gain} energy $\sim -0.9t_h$ for each holon at
quarter-filling$.$ This gaining of the kinetic energy is in contrast to the
energy \emph{cost} for a hole-dipole escaping away from the stripe: $%
\mathcal{E}^d\sim J_s$ according to Eq.(\ref{edipole}) \cite{remark1}. The
energy difference will stabilize such a \emph{metallic} stripe of
quarter-filling, at least in a local minimal sense. In literature, a
metallic stripe based on the $t-J_z$ model has been discussed in Ref. \cite
{neto1}. The spectral density for a single hole in the stripe phase has been
also studied in Ref. \cite{eder1}.

Based on the phase string picture, a metallic stripe discussed above can be
regarded as a line which connects those branch-cuts between the starting and
ending points of holons in Fig. 3. As emphasized before, a phase string is
so singular that it can only be eliminated when one holon tightly tails
another, which leads to a 1D line of charges rather than a pairing of holes.
Physically it is the phase string effect that ultimately threads holes
together in a metallic stripe.

As a natural extrapolation, one may picture a stripe liquid as composed of
mesoscale stripes threaded by phase strings, after the melting of static
stripes. Eventually, say at high enough temperature, the most elementary
entities, i.e., hole-dipoles, should be also present, if spins are still
ordered (Bose condensed). Experimentally, how to detect such mesoscale
stripes or even individual hole-dipole objects is an interesting subject to
further explore in the present framework.

\emph{Diagonal stripe. }For a Type B dipole, its minimal dipole moment is
shown in Fig. 4(b). Here one pole is located inside a plaquette while the
pole of the holon is at a corner of the plaquette. Fig. 5(b) shows a natural
stripe configuration composed of the minimal Type B dipoles in Fig. 4(b). It
is a stripe along the diagonal direction, which remains as an anti-phase
domain as the argument leading to Eq. (\ref{deltaphi}) still holds here. But
it becomes an insulating stripe instead, as holons cannot directly move
along the stripe.

It has been argued that hole-dipoles of Type B may be more stable at low
doping as it costs less superexchange energy. By forming a diagonal stripe,
there is no obvious gain in the hole's kinetic energy, except for some
virtual process. In a single hole-dipole case, such a virtual process may be
understood as each holon continuously hops around and then back to its
origin, leaving irreparable phase strings on the paths which result in a
spin-dipole configuration. When these holes are lined up in the stripe
configuration, the phase strings are canceled out each other during the
virtual hopping process such that there is no spin twist left behind away
from the stripe [Fig. 5]. So for both types of stripe in Fig. 5, the healing
of the spin AF order away from the domain wall indicates an effective \emph{%
kinetic energy gain} for virtual hole hopping out of the domain wall.
Furthermore, holons gains \emph{additional} \emph{kinetic energy} in the
metallic stripe, but the \emph{superexchange energy} of the domain wall
seems favored in the diagonal stripe, under a given number of holes.

One may envisage some other kinds of static stripes. A detailed analysis of
which type of stripes is most energetically favorable in a given doping
concentration is beyond the scope of the present paper, and will be explored
elsewhere. Finally, we note that various charge orderings due to hole-hole
long-range (dipolar and Coulomb) interactions have been also studied
numerically in Ref. \cite{stojkovic}, based on a spin-density-wave, or
itinerant approach. Our strong-couling (t-J type model) approach seems to
provide a consistent picture concerning the formation of static stripes,
though many details still need a further comparison.

\section{Discussions}

In this paper, we have studied how holes behave in a spin ordered background
at low doping. In a doped AF Mott insulator described by a derivative model
of the $t-J$ Hamiltonian, known as the phase string model, the kinetic
energy of doped holes gets strongly frustrated by the phase string effect.
These holes induce nonlocal effects in the spin background and behave like
dipoles. In contrast to the hole dipole picture of Shraiman and Siggia \cite
{ss} based on the semiclassical treatment, a holon constitutes one of the
two poles in the present configuration instead of sitting at the center of
the dipole \cite{ss}. Consequently, hole dipoles are localized or
self-trapped objects in space rather than mobile ones in the
Shraiman-Siggia's case. Such a localization or self-trapping of a hole can
be intuitively understood directly in terms of the irreparable phase string
effect \cite{1hole}.

That the hole kinetic energy is strongly suppressed at low doping in the
phase string model is in contrast to the slave-boson theories \cite{fRVB} of
the $t-J$ model. In the latter cases, holes are relatively free moving on
various RVB backgrounds. The difference arises from the fact that in the
bosonic RVB background \cite{lda}, short-range\emph{\ AF correlations} are
much stronger than in fermionic RVB theories, which are highly unfavorable
to the hole motion.

Due to the frustrations of the hole kinetic energy, the dipole-dipole
interaction dominates the low-energy, long-wavelength physics. Holes
correlate with each other through such an interaction which, instead of
leading to a real-space pairing of holes, causes an inhomogeneous
instability -- stripe instability with hole-dipoles collapsing into a 1D
line. An anti-phase, quarter-filling metallic stripe can be regarded as a
pile up of Type A dipoles, while an anti-phase, diagonal insulating stripe
may be naturally considered as composed of Type B dipoles. Here the phase
string effect provides the basic thread for forming stripes: holes tightly
follow each other's paths in order to minimize the frustrations of the phase
string effect and to release the kinetic energy.

Theoretically, stripes and stripes related issues have been studied
intensively in literature \cite{review1}, motivated by experimental
discovery \cite{tran} of static stripes in curpates. In fact, insulating
stripes as a mean-field solution was first obtained \cite
{zaanen,schulz,machida} in the Hubbard and $t-J$ models at low-doping even
before the experiment. The present approach has gone beyond the mean-field
approximation in dealing with the motion of doped holes in an
antiferromagnet. It reveals that as a ``renormalization'' effect there does
exist a \emph{residual} long-range interaction for holes to ``talk'' to each
other and to collectively form stripes. This approach provides a framework
to further systematically investigate various properties associated with
stripe phenomenon. Finally, in literature stripes as composed of
ferromagnetic bonds beyond the $t-J$ type model have been previously
proposed \cite{gooding}. But the origin of the underlying physics as well as
the consequence are different.

Some interesting questions remain to be answered. Firstly, if static stripes
melt, are there dynamic stripes present at higher energy or higher
temperature? If they exist, it is also a very challenge problem to
mathematically describe such dynamic stripes of ``finite sizes''. Recently a
Z$_2$ gauge theory of ``stripe fractionalization'' has been proposed by
Nussinov and Zaanen \cite{nussinov} based on a ``topological'' order of the
sublattice parity. The Z$_2$ phase string depicted in Fig. 1(a) indeed
originates from the \emph{change} of the sublattice parity of spins due to
the hole hopping. But the phase string model of Eqs. (\ref{hh}) and (\ref{hs}%
) is not simply a Z$_2$ gauge theory because the original building blocks
are \emph{holons }and\emph{\ spinons}, not in a stripe fractionalization
description.

Of course, in the end the holons are actually confined in the spinon
condensed phase and elementary objects become hole-dipoles. A mesoscale
stripe structure may be expected as remaining clusters at high energy or
temperature, threaded by phase strings, which should eventually break into
hole-dipoles. But we note that the above dipole picture and stripe
instability are all discussed when spinons are Bose condensed. When such a
condensation is gone at high temperature or larger doping, the dipole and
stripe pictures are no longer meaningful in this framework.

Secondly, what is the relation between superconductivity and stripe phases?
So far we have not explored the possibility of superconducting condensation
in the spinon condensed phase, where stripes present. In the phase string
model, superconducting condensation happens when holons are Bose condensed 
\cite{string1,muthu}. It can occur at higher doping where no spinon
condensation is present \cite{muthu}. Thus, superconducting phase and stripe
phase seem not necessarily concomitant in this framework. Nevertheless, the
origin of superconductivity and that of stripes can all be attributed to 
\emph{kinetic-energy-driven mechanisms}. Since the phase string effect plays
the concrete role of kinetic-energy frustrations, various low-temperature
phases including stripes, superconductivity, and even the pseudogap
phenomenon may be regarded as different ways in reducing kinetic energies in
such a system \cite{kou}.

\acknowledgements
We acknowledge stimulating discussions with N. P. Ong which partially
motivated the present work. This work is partially supported by NSFC grant
nos. 90103021, 10247002, and 10204004.


\begin{thebibliography}{99}
\bibitem{remark1}  Note that the self-energy of a hole-dipole also includes
a hopping energy as the holon confined to the dipole may still hop around
locally according to $H_h$ [Eq.(\ref{hh})]. Rigorously speaking, one needs
to consider both hopping energy and the confining potential energy Eq.(\ref
{edipole}) in determining the stability.
\end{thebibliography}
\end{document}